\documentclass[letterpaper]{article}
\usepackage{aaai17_withoutcpr,times,helvet,courier,xspace}
\usepackage{amsmath,amsfonts,tabularx,subcaption,multirow,graphicx,longtable,microtype,booktabs}
\usepackage[hyphens]{url}
\usepackage[table,xcdraw]{xcolor}
\usepackage{caption,subcaption}
\usepackage[square,sort,comma,numbers]{natbib}

\newcommand{\para}[1]{{\vspace{4pt} \bf \noindent #1 \hspace{10pt}}}

\newcommand{\eg}{e.g.,\ }

\begin{document}

\title{Why Do Men Get More Attention? \\Exploring Factors Behind Success in an Online Design Community}

\author{Johannes Wachs$^{\rm\dagger}$ \hskip0.45cm Anik{\'o} Hann{\'a}k$^{\rm\dagger}$ \hskip0.45cm Andr{\'a}s V{\"o}r{\"o}s$^{\rm *}$ \hskip0.45cm B{\'a}lint Dar{\'o}czy$^{\rm\ddagger}$ \\
	\\
	\small{$^{\rm\dagger}$Center for Network Science, Central European University \hskip1cm $^{\rm *}$Chair of Social Networks, ETH Zurich, Zurich, Switzerland} \\ \small{$^{\rm\ddagger}$Institute for Computer Science and Control (MTA SZTAKI), Hungarian Academy of Sciences}}

\maketitle
\date{}
\begin{abstract}

Online platforms are an increasingly popular tool for people to produce, promote or sell their work. However recent studies indicate that social disparities and biases present in the real world might transfer to online platforms and could be exacerbated by seemingly harmless design choices on the site (\eg recommendation systems or publicly visible success measures). 
In this paper we analyze an exclusive online community of teams of design professionals called Dribbble and investigate apparent differences in outcomes by gender. Overall, we find that men produce more work, and are able to show it to a larger audience thus receiving more likes. Some of this effect can be explained by the fact that women have different skills and design different images. Most importantly however, women and men position themselves differently in the Dribbble community. Our investigation of users' position in the social network shows that women have more clustered and gender homophilous following relations, which leads them to have smaller and more closely knit social networks. Overall, our study demonstrates that looking behind the apparent patterns of gender inequalities in online markets with the help of social networks and product differentiation helps us to better understand gender differences in success and failure.


\end{abstract}

\section{Introduction}

Research in the social sciences has shown that both individual and social network attributes impact individual success in education, the workplace, and the job market. However, the specific mechanisms enabling or hindering success are highly dependent on social context: the available channels for social contact, the constraints on social ties, the channels for social influence, group sizes, and other factors clearly influence individual success. 

The recent growth in popularity of online platforms for social interactions (e.g. Twitter, Instagram) and job search (e.g. Linkedin, freelancer.com) changes the social mechanisms that determine individual success. Currently, we know very little about how inequalities emerge in these new types of communities. We do know, however, that the design of the sites matters: some researchers express concerns that the use of algorithms and public feedback might retain or even reinforce inequalities in success based on, especially, race and gender~\cite{lustig2016algorithmic,lee-chi15}. Empirical work in online freelance communities \cite{teodoro-2014-cscw,thebault-2015-cscw,hannak-2017-cscw} and on collaboration in teams~\cite{vasilescu2013gender,de2015game} also highlight the presence of gender inequalities.  

Some of these communities combine the open nature of online social networks with the professional aspects of real-world labor markets. Users invest in their identities by showing work, exchanging ideas, and collaborating in visible ways. Over time and with great investment users shape permanent identities with reputations and social capital, just like in the real world. Because of the online nature these identities operate in a different social environment. Online ties between people are about sharing access and have much lower costs, and exist at much larger scales. At the same time online platforms shape social interaction are highly structured and governed by algorithms. Our goal is to explore the inequalities that emerge from this combination of a scaled-up social environment with highly structured systems.

In this paper we analyze Dribbble, the most ``elite'' online community for digital and graphic designers. The site allows designers to showcase their work in web design, illustration, and other creative areas, follow artists whose work they appreciate, discuss design ideas, and work on collaborative projects in teams. Dribbble enjoys high prestige in the worldwide community of digital and graphic designers, as it is invitation-only and provides a good platform for advertising one's work. We crawled the pages of all 994 teams on the site, 6,215 users involved in one of the teams, and finally all 60,406 images created by these teams.

Our questions are \textit{do men and women have different success rates on Dribbble?}, and if yes, \textit{what are the factors contributing to the differences?} We separately analyze the effects of individual user characteristics, activity on the site, production patterns, and social network structure to understand how much each of these factors contributes to success of individuals and where gender differences may be reinforced. 

Using the variables extracted from our data set, we define three measures of user success: the average number of views, likes, and responses the works of a user receive. Using regression analysis, we establish that men are more successful according to all measures, even after controlling for basic individual characteristics extracted from the profile information.

\begin{figure*}[t]
	\centering \includegraphics[width=0.98\textwidth]
	{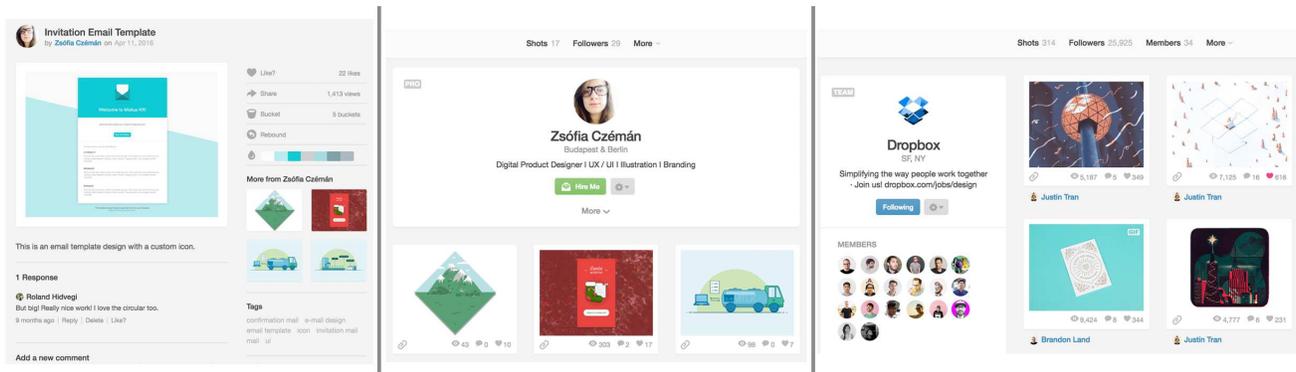}
	\vspace{-1em}
	\caption{Shot, User and Team Pages on Dribbble.}
	\label{fig:shot_user}
\end{figure*}

Since skills and social background might determine the kind of work people produce, we next investigate whether this is true on Dribbble and how much of the gender differences can be explained by such factors. We create a measure for skill and image ``genderness'' using data mining techniques. Interestingly, even though skills are not strongly divisive, our classifier can predict with 0.72 AUC if an image was created by a man or a woman. This suggests that indeed some men and women are creating different art. Once controlling for these variables in our models we find that the relationship between gender and outcome is no longer significant. 

Finally, we investigate social behavior and network effects on the social network underlying Dribbble. We run Exponential Random Graph Models (ERGM) and the results suggest that women have fewer ties but more cohesive social networks than men. When we return to our original model of success and control for these network features, we find again that gender no longer has a significant impact on success.

\section{Related Work}

What determines whether people succeed at school, in the job market, or in the workplace? Social scientists recognized long ago that individual characteristics, especially those related to socio-economic background and behavior are important predictors of performance and achievement \cite{angristlang2004,curcio2006sleep}. However, how such individual factors determine success greatly depends on social context and especially on the network of social relations between people \cite{coleman1994}.

A long line of social research from various disciplines demonstrates that individual outcomes and social networks are dependent on one another and they are shaped by a couple of social mechanisms that are observable across various settings, both online and offline. These include homophily in characteristics such as sex, race, ethnicity or family background \cite{feld1982,mcphersonetal2001}, triadic closure \cite{heider1946,cartwrightharary1956}, clustering and hierarchy formation \cite{davis1970}, and social influence on individual attitudes and behavior \cite{veenstradijkstra2011,marsdenfriedkin1993,turner1991}.

Some studies specifically show that people's position in informal social networks can affect their performance. For example, recent research on peer influence and social networks among academics has found that scientific collaboration impacts academic success \cite{petersen2014reputation,petersen2015}. On the macro level, the centrality of individuals in collaboration networks is positively associated with their success \cite{Sarigoel2014,servia2015evolution}.

Thus, in general it seems that having many connections and a cohesive network may promote individual success. This may especially be true in cases when success is closely linked to informal social status: when people can use or mobilize their social relations to ``generate success'', such as popularity or expressed appreciation -- this is exactly the case in the empirical study described in this paper.

Informal social networks and individual outcomes co-evolve through selection and influence processes \cite{veenstradijkstra2011}: people select the peers they associate with not independently of individual characteristics (e.g. homophily); in turn, friends, role models, or groups influence how people behave and perform. This links the question of success closely to the emergence of inequalities in social groups. Many studies focus on gender-based inequalities in education systems~\cite{jacobs1996gender}, hiring~\cite{pager-2008-sociology,clauset2015systematic}, scientific careers~\cite{lee2013bias}, or work contexts. An important limitation of many empirical studies in the presented research line is the context dependence of their results. While most studies agree that women (and minorities in general) have worse chances of succeeding, the factors highly depend on the community being studied.

The ongoing migration of both professional and social life to to online platforms changes the mechanisms related to success and inequalities~\cite{sandvig2014auditing,robinson2015digital}. Demographic characteristics and status signals are visible on users' online profiles~\cite{tang_dasfaa11} while the pathways to success largely depend on website design and invisible algorithms~\cite{lee-2015-chi}. Studies in online social networks find that influence propagates through connections with more trust and more status~\cite{adali2010measuring,hannak2014get,munger2016tweetment,ajrouch2005social}. 

\begin{table*}[t]
	\centering
	\begin{tabular}{ll}
		\toprule
		Shot level & Creation Date, Creating Team/User, Shot, Shot Tags, \# of Views, \# of Likes, \# of Responses \\
		\midrule
		User level & Name, Bio, Skills, Team, Premium Account, \# of Shots, \# of Followers \\
		\midrule
		Team level & Members, \# of Shots, \# of Followers \\
		\bottomrule
	\end{tabular}
	\caption{Extracted features from Dribbble.}
	\label{tab:features}
\end{table*}

A few recent studies explore success and inequalities in online labor markets~\cite{hannak-2017-cscw,thebault-2015-cscw,NBERw22776} and find that the new mechanisms such as public review systems and algorithmic search might amplify inequalities~\cite{Fradkin:2015:BRO:2764468.2764528,pan-2007-trust}. In these settings however there is no clearly measurable underlying social network, and thus the relevance of social effects cannot be easily assessed. Finally, works investigating collaborative team-based platforms such as github~\cite{vasilescu2013gender} or online video games~\cite{de2015game} capture the complex interaction of individual and group success.

\section{Data and Extracted Features}

\paragraph{Dribbble}
Dribbble, founded in 2009, is an exclusive community for showcasing user-made artwork in graphic design, web design, illustration, photography, and other creative areas. It has an Alexa rank of 1,012 (as of 01-05-2017) making it the second most visited online community for work in digital design after Behance. In contrast with Behance, Dribbble has an invite-only membership system. Users can only upload their work after receiving an invite from a current member. Moreover, users can only post 48 images in a month and five shots in a day.  The result is a high standard of work on the platform and the sense of belonging to an ``elite'' community. Moreover, Dribbble is a true community in the sense that nearly all users share their identities by linking to their social media accounts and personal home pages, and by uploading photo portraits. Dribbble facilitates job matching between companies and users paying for premium accounts. A significant amount of users pay for this service, highlighting Dribbble's structural importance in this field.

The service has become a platform for some of the most abstract and compelling parts of the creative process. Users can see what their peers are creating and both solicit and give feedback. The site facilitates active use by allowing users to view, like, and comment on work, and to follow others. There is a special subscription that allows organizations to show their work together. In fact many leading IT companies and design boutiques are represented as a ``team'' on the site. Everything on Dribbble is basketball themed: \emph{players} or \emph{teams} post the art work which they refer to as \emph{shots}. We will follow this convention throughout the paper.

\paragraph{Data Collection}
Our data collection focuses on three levels of the site: teams, users, and shots. As shown in Figure~\ref{fig:shot_user} teams, users, and shots each have an associated profile page. We first crawl the pages of all 994 teams on the site, then we extract the pages all 6,215 users who made a shot while on a team, and finally we collect all 60,406 shots created by teams. About three-fourths of all shots have an individual user authorship tag in addition to the team authorship. All the crawling was done between September and November 2016. During our crawl, we made sure to respect the robots.txt and impose minimal load on Dribbble; we sent a maximum of 1,440 requests per day. 

Besides their Dribbble profiles we also collect data about users and teams from their linked Twitter accounts. 86\% of teams and 70\% of users have twitter accounts. Since it is not possible to obtain the Dribbble follower network directly, we later use Twitter data as a proxy for the social network underlying Dribbble.

\paragraph{Extracted features}
Table~\ref{tab:features} shows the variables we extract from the data set we collected. The three columns show the three types of pages we crawl. For each object (shot, user, team) we have descriptive variables, such as creation date of a shot, listed skills of a user, or the size of a team. We also extract dynamic features such as number of views, likes, and comments, which we use as measures of success or popularity.

\begin{figure*}[t!]
	\centering
	\begin{subfigure}[t]{0.33\textwidth}
		\includegraphics[width=1\textwidth]{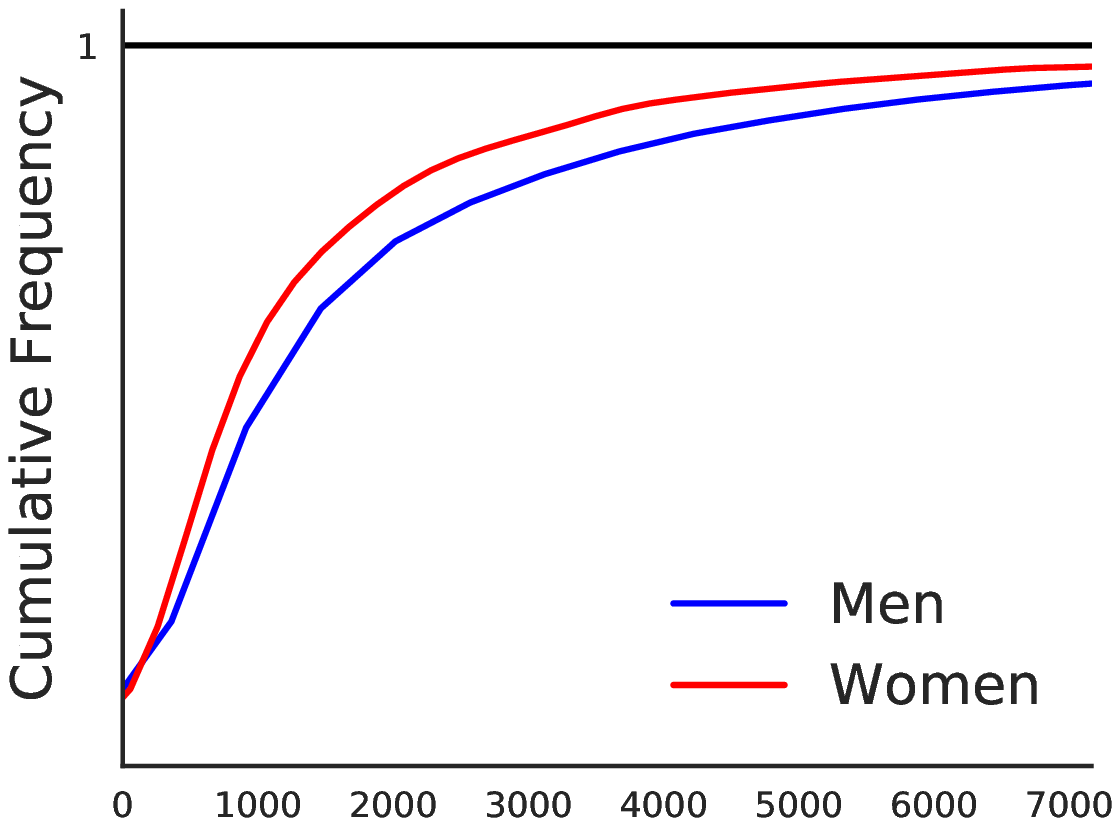}
		\caption{User Average Views}
		\label{fig:views_summary}
	\end{subfigure}
	\begin{subfigure}[t]{0.33\textwidth}
		\includegraphics[width=1\textwidth]{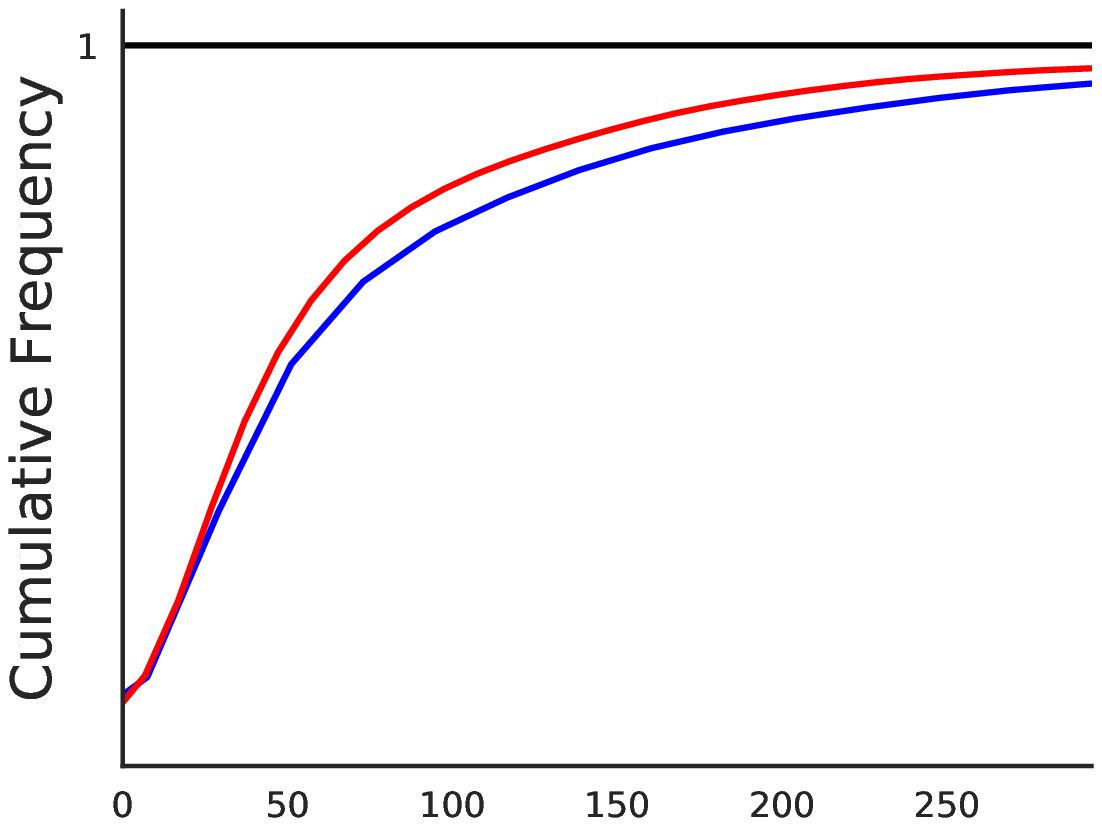}
		\caption{User Average Likes}
		\label{fig:likes_summary}
	\end{subfigure}
	\begin{subfigure}[t]{0.33\textwidth}
		\includegraphics[width=1\textwidth]{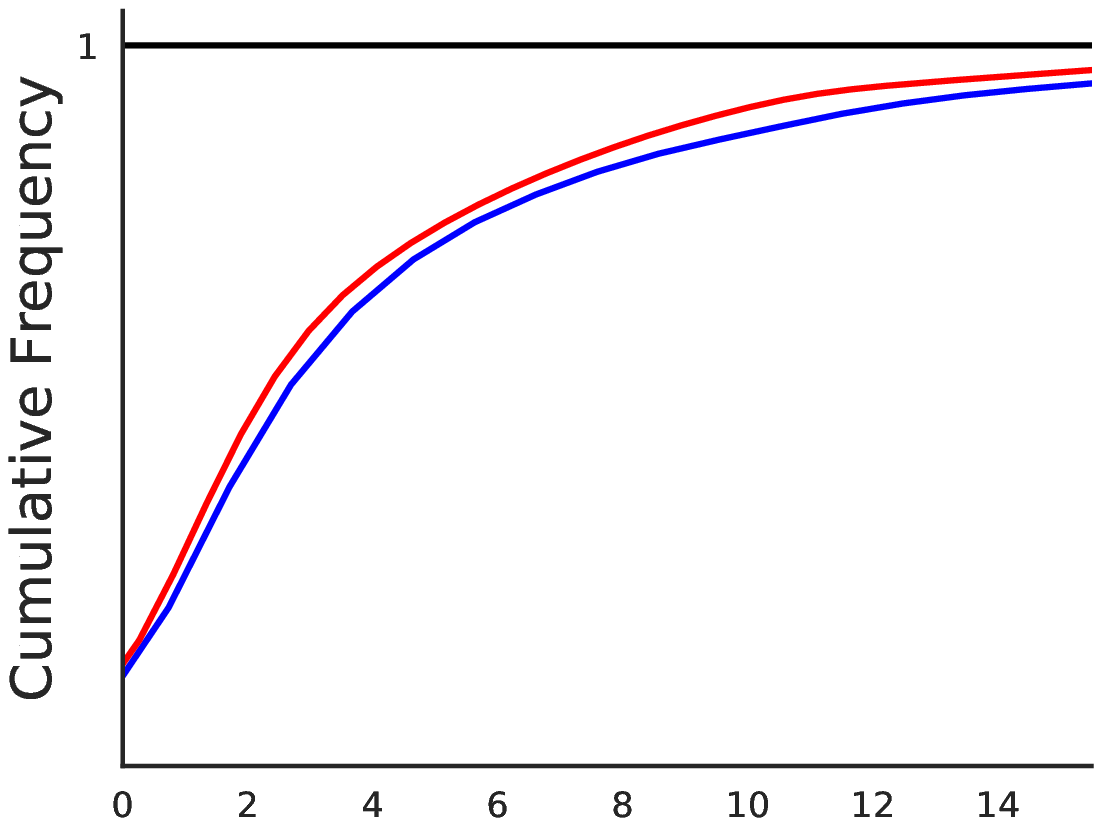}
		\caption{User Average Responses}
		\label{fig:responses_summary}
	\end{subfigure}
	\caption{Cumulative distribution functions for user average views, likes, and responses by gender. \textit{(Note: CDF is cut off at 95\% of the global average.)}}
	\label{fig:outcome_1}
\end{figure*}

\para{Inferring gender.}
Since the profiles do not directly list gender, we infer them from the users' first names using the US baby name~\cite{babyname} data set. This is commonly and successfully used method in many studies~\cite{ruths-aaai13,tang_dasfaa11,karimi-www16,cfpd_race,fiscella-hsr06}. We are especially confident that users give their real names because of the site's nature as a platform for designers to advertise their work and gain a following. Our first-pass inference of gender is the probability of being male to each candidate's name based on its occurrence in the name data set as male. For any user with a name not in the database or an ambiguous gender score (i.e. greater than 10\% and less than 90\%) we manually check their self-portrait on Dribbble and on linked social media account. We were able to classify all but 77 out of 6,215 users as male or female with high confidence (in the rest of the paper we drop these 77 users). Our final dataset contains \emph{4654 males} and \emph{1484 females}.

\section{Results}
\label{sec:res}

In this section, we investigate the relationship between users' success and popularity on Dribbble and their gender. Since we find that gender differences in success do exist, we dig deeper in search of user characteristics or social factors which might explain them. Specifically, we focus on two kinds of potential explanations. First, we define various measures that capture how users create products and investigate how these variables affect success and how they relate to gender. Second, we explore the impact of the social aspect of Dribbble by identifying measures of social behavior and network position.

\subsection{Gender Differences in Outcomes}

Dribbble provides various ways for users to express appreciation for each other's work, such as liking a shot or leaving a response. These markers of popularity are shown on each product's page. We can see how many people viewed a shot (this is purely a matter of clicking when the thumbnail shows up in the audience's feed or following a direct link). We see the number of likes, which is a clear signal of appreciation. Finally, users sometimes comment on an image. The number of responses is a stronger indication of interest, since leaving a public feedback requires more trust or engagement. Although not all comments are positive, we argue they are likely positive in general and that they are a success signal when aggregated on the user profile. We also note that they are non-anonymous. As we want to measure user success, we aggregate these values for all shots created by an individual and use the average view, like, and response counts as three distinct success measures. 

\begin{figure*}[t!]
	\centering
	
	\begin{subfigure}[b]{0.31\textwidth}
		\includegraphics[height=3.5cm]{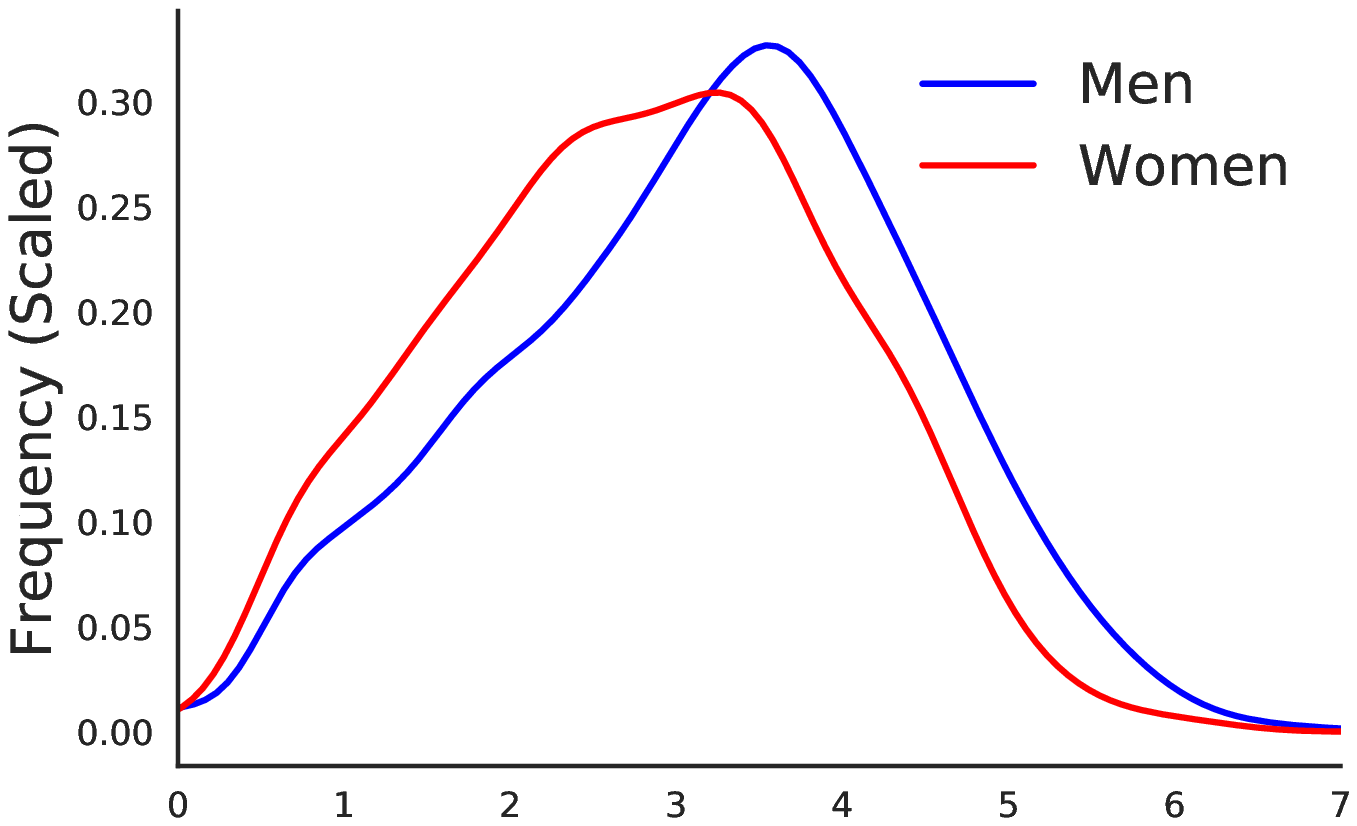}
		\caption{Distribution of the log of shot count of users by gender.}
		\label{fig:views_summary}
	\end{subfigure}
	\begin{subfigure}[b]{0.31\textwidth}
		\includegraphics[height=3.5cm]{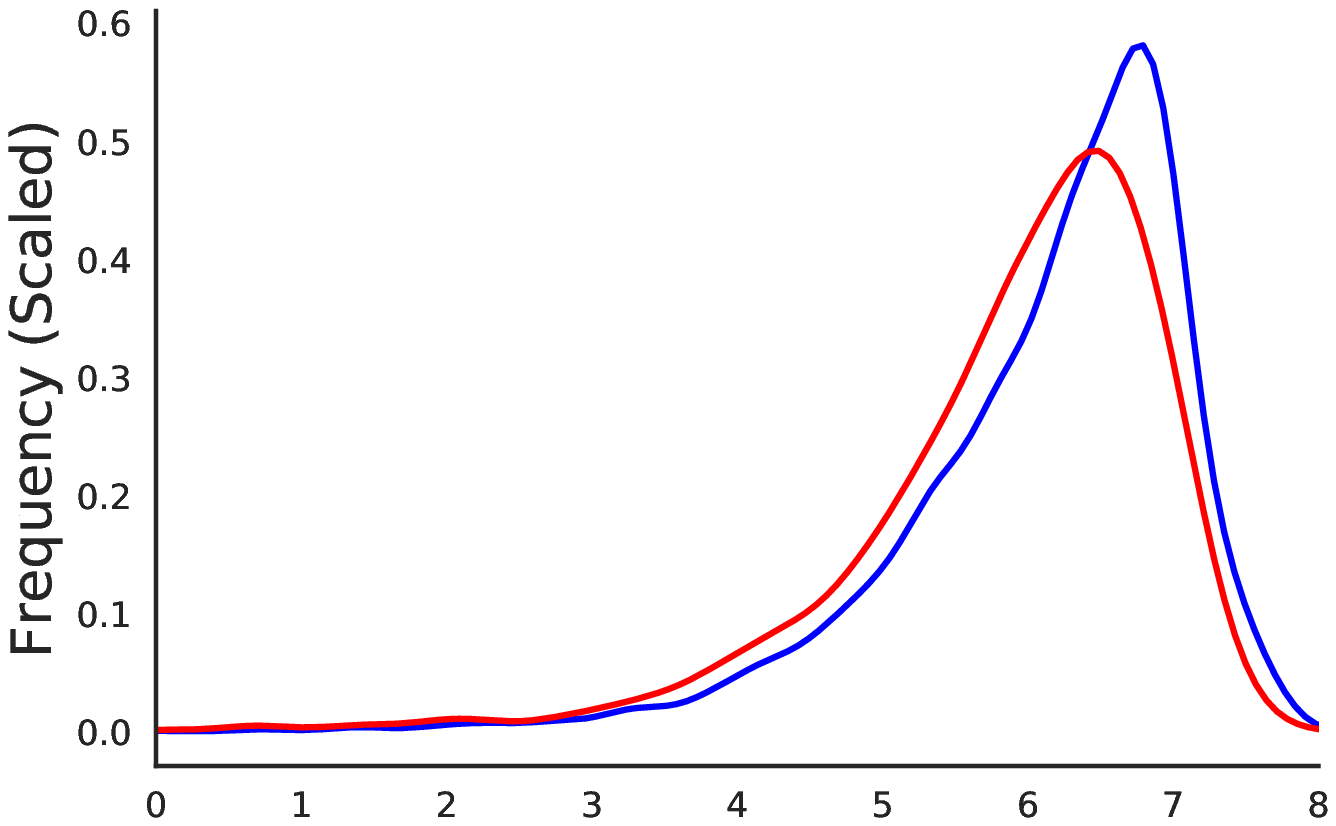}
		\caption{Distribution of the age of user accounts in log days by gender.}
		\label{fig:likes_summary}
	\end{subfigure}
	\begin{subfigure}[b]{0.31\textwidth}
		\includegraphics[height=3.5cm]{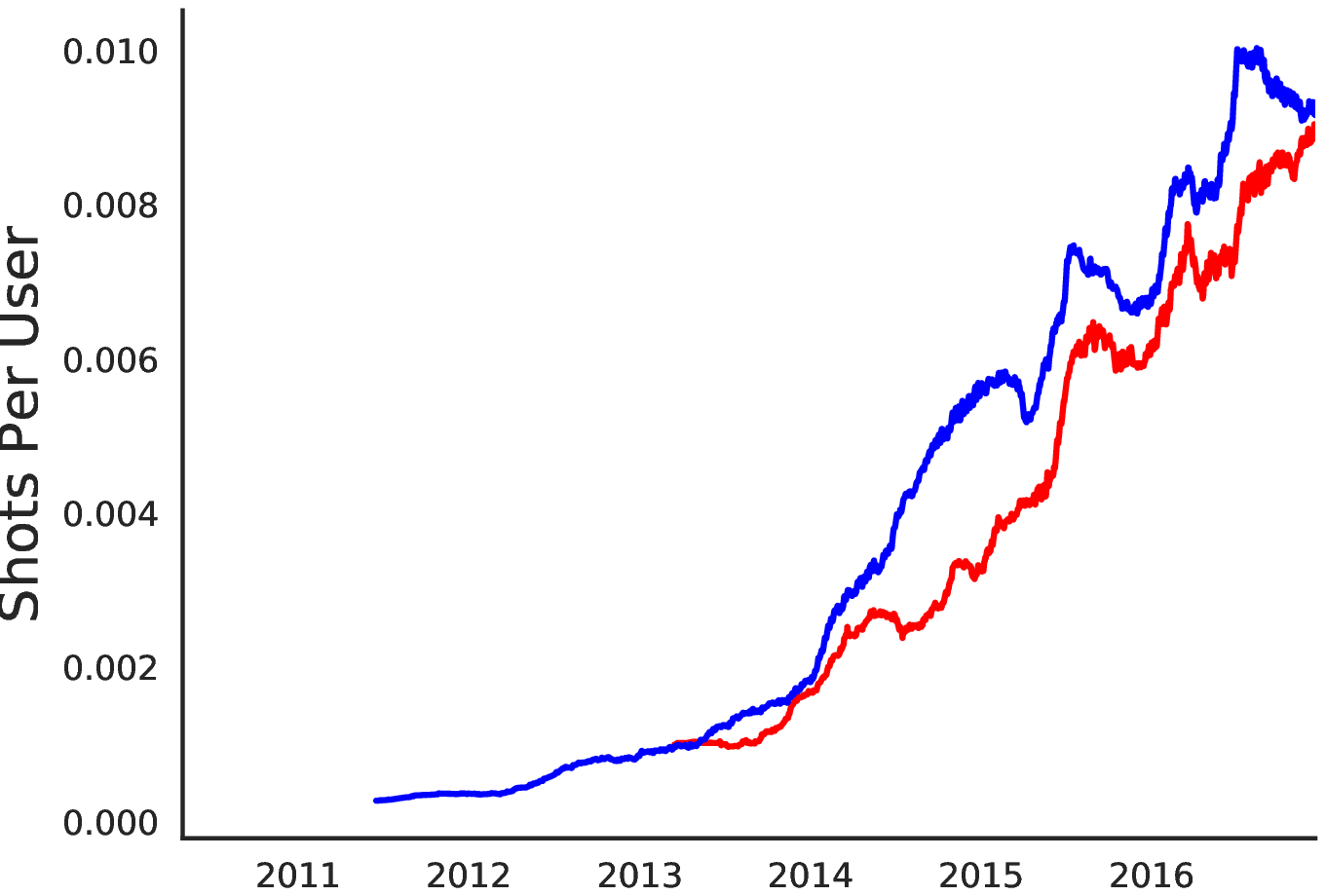}
		\caption{Shots per day per person over time, 90-day moving average by gender.}
		\label{fig:responses_summary}
	\end{subfigure}
	\caption{Summary statistics comparing male and female activity on the site.}
	\label{fig:outcome_2}
\end{figure*}

At a first glance, women and men have different success rates according to all three measures in this community. Figure~\ref{fig:outcome_1} shows the gender distributions for the log of the three success measures. (The distribution has a long tail, with only men at very high success rates. For easier interpretation, we cut the CDF at 5\% of the global average on either side of the plot.) Men have more views, likes, and responses on average. We find the differences to be significant ($p \leq 0.01$) when applying a Kolmogorov-Smirnov test to compare the distributions. The common language effect sizes between men and women, calculated using the Vargha-Delanay $A$-measure, are .57, .54, and .54 for views, likes, and responses, respectively. Though the effect sizes are relatively small, we see below that they remain statistically significant even after adding multiple controls.

Considering the user characteristics that could potentially explain success differences, we find that men and women have different levels of participation on Dribbble over time. As shown in Figure~\ref{fig:outcome_2}, male accounts are on average 24\% older and post 60\% more shots than female accounts. We can also see that until very recently men posted new work more frequently than women.

Having a premium Dribbble account and identifying as a leader apparently also covary with gender and may be related to success. 38\% of men pay for the premium account service or \textit{Pro Badge} compared to 23\% of women. We compared user bios based on the self-descriptions of individuals: processed bios containing the strings ``founder", ``director", ``manager", or ``partner" were labeled as \textit{self-described leaders}. We find that 19\% of men and 10\% of women are self-described leaders. There is no significant correlation between having a pro-badge and self-describing as a leader for men or women, while at the same time both variables correlate with higher output.

Arguably, the size of the team a user is part of may also influence his or her success on Dribbble. This is not only because large companies are present with larger teams on the site, but also because of the larger potential for  \textit{collegial interactions} and feedback present in larger teams. Although we do not find a statistically significant difference between the average team size of women (mean size = 9.5) and men (mean size = 10), we include it as a control in our models.

We first examine if the described differences in production, incumbency, investment, status and collegial interactions can explain the differences in our success variables using linear regressions. Specifically, we predict the effect of gender on the logarithm of a user's average views, likes, and responses while controlling for:

\begin{itemize}
	\item log of number of shots by the user (productivity), 
	\item log of days since first shot of the user (incumbency),
	\item whether the user has a pro-badge (investment),
	\item whether the user self-describes as a leader (status),
	\item log of the size of team (collegial interactions).
\end{itemize}

Table~\ref{tab:success_base} shows the results for our models setting the {\it number of views, number of likes and number of responses} as dependent variable, respectively. From the first column, we see that being experienced, having a Pro Badge, and having a large team all positively contribute to the average number of views a user receives on their shots. However even after controlling for these variables, males have significantly more views. 
Next, we look at the second column which shows results for the average number of likes as the dependent variable. The results look fairly similar, except that having joined the site earlier relates to having more likes. Moreover, being male, while still significant, has a smaller positive effect than in case of the views. Finally, we see a slightly smaller effect of being male for the number of responses. We note that while we find that most of our user variables correlate with success, these models do not predict success very well, achieving low $R^{2}$ values. 

With our three initial regression models, we have established that there are some robust differences in individual success between men and women. Men receive more views, likes, and responses. The main takeaway from this analysis is that while we see differences between men and women in different measures of success, these are not explained by basic user-level characteristics. In the following sections we investigate two social factors that might contribute to the observed gender discrepancies in success: that {\em men and women are creating different products} and that {\em men and women have different social network structures}.

\begin{table}[t!]
	\begin{centering}
		\scriptsize
		\begin{tabular}{lr@{}lr@{}lr@{}l} 
			\toprule
			User Averages (log): & \multicolumn{2}{c}{Views} & \multicolumn{2}{c}{Likes}  & \multicolumn{2}{c}{Responses} \\
			\midrule
			Log(Number of Shots) & $\mathbf{0.09}$ & $^{***}$ & $\mathbf{0.11}$ & $^{***}$ & $-0.004$&\\
			Log(Age of Account) & $0.03$ &  & $\mathbf{-0.15}$ & $^{***}$ & $\mathbf{0.04}$ & $^{**}$\\
			Leadership Word in Bio & $0.05$ &  & $0.08$ & & $\mathbf{0.08}$ & $^{*}$\\
			Pro-badge & $\mathbf{0.49}$ & $^{***}$ & $\mathbf{0.44}$ & $^{***}$ & $\mathbf{0.29}$ & $^{***}$\\
			Log(Team Size) & $\mathbf{0.39}$ & $^{***}$ & $\mathbf{0.21}$ & $^{***}$ & $\mathbf{0.002}$ & $^{***}$\\
			Is Male & $\mathbf{0.19}$ & $^{***}$ & $\mathbf{0.10}$ & $^{*}$ & $\mathbf{0.07}$ & $^{**}$\\
			\midrule
			
			$R^{2}$ & $\mathbf{0.13}$ &  & $\mathbf{0.12}$ & & $\mathbf{0.06}$ & \\
			
			\bottomrule
		\end{tabular}
		\caption{OLS regressions on user-level variables to predict impact of gender on success. (\textit{Note:}  $^{*}p < 0.05$; $^{**}p < 0.01$; $^{***}p < 0.001$)} 
		\label{tab:success_base}
	\end{centering}
\end{table}

\begin{table}[t]
	\centering
	\scriptsize
	\begin{tabular}{llll}
		\toprule
		\textbf{Most Male Skills}& \textbf{Most Female Skills} \\ \hline
		interfaces   & calligraphy  \\
		productmanagement     & copywriting  \\
		objectivec     & research  \\
		iosdev     & information  \\
		compositing     & handlettering  \\
		framer     & socialmedia  \\
		gui     & visualcommunication  \\
		ruby     & gamedesign  \\
		apparel     & branddev  \\
		illustrator     & drawing  \\
		identity     & ecommerce  \\
		\bottomrule
	\end{tabular}
	\caption{Ten most male and female self-reported skills.}
	\label{tab:skills_common}
\end{table}

\begin{table*}[t!]
	\centering
	\scriptsize
	\begin{subtable}[t!]{0.48\textwidth}
		\centering
		\begin{tabular}{lr@{}lr@{}lr@{}l} 
			\toprule
			User Averages (log): & \multicolumn{2}{c}{Views} & \multicolumn{2}{c}{Likes}  & \multicolumn{2}{c}{Responses} \\
			\midrule
			Log(Number of Shots) & $\mathbf{0.08}$ & $^{***}$ & $\mathbf{0.11}$ & $^{***}$ & $-0.002$&\\
			Log(Age of Account) & $0.02$ &  & $\mathbf{-0.14}$ & $^{***}$ & $\mathbf{0.04}$ & $^{**}$\\
			Leadership Word in Bio & $0.05$ &  & $0.08$ & & $\mathbf{0.08}$ & $^{*}$\\
			Pro-badge & $\mathbf{0.45}$ & $^{***}$ & $\mathbf{0.43}$ & $^{***}$ & $\mathbf{0.28}$ & $^{***}$\\
			Log(Team Size) & $\mathbf{0.38}$ & $^{***}$ & $\mathbf{0.32}$ & $^{***}$ & $\mathbf{0.21}$ & $^{***}$\\
			Is Male & $\mathbf{0.14}$ & $^{***}$ & 0.08 &  & 0.06 & \\
			User Lists Male Skill& $\mathbf{0.21}$ & $^{***}$ & 0.04 & & 0.04 & \\
			User Lists Female Skill & $\mathbf{-0.25}$ & $^{***}$ & $\mathbf{-0.16}$ & $^{***}$ & $\mathbf{-0.12}$ & $^{*}$\\
			
			\midrule
			
			$R^{2}$ & $\mathbf{0.14}$ &  & $\mathbf{0.13}$ & & $\mathbf{0.06}$ & \\
			
			\midrule
		\end{tabular}
		\caption{The effects of user skill ``genderness''.}
	\end{subtable}
	\hfill
	\begin{subtable}[t!]{0.48\textwidth}
		\centering
		\begin{tabular}{lr@{}lr@{}lr@{}l} 
			\toprule
			User Averages (log): & \multicolumn{2}{c}{Views} & \multicolumn{2}{c}{Likes}  & \multicolumn{2}{c}{Responses} \\
			\midrule
			Log(Number of Shots) & $\mathbf{0.11}$ & $^{***}$ & $\mathbf{0.34}$ & $^{***}$ & $-0.004$&\\
			Log(Age of Account) & $0.03$ &  & $\mathbf{-0.11}$ & $^{***}$ & $\mathbf{0.04}$ & $^{**}$\\
			Leadership Word in Bio & $0.06$ &  & $0.09$ & & $\mathbf{0.08}$ & $^{*}$\\
			Pro-badge & $\mathbf{0.47}$ & $^{***}$ & $\mathbf{0.42}$ & $^{***}$ & $\mathbf{0.27}$ & $^{***}$\\
			Log(Team Size) & $\mathbf{0.40}$ & $^{***}$ & $\mathbf{0.34}$ & $^{***}$ & $\mathbf{0.23}$ & $^{***}$\\
			Is Male & 0.06 &  & 0.02 &  & 0.02 & \\
			Shot Maleness & $\mathbf{1.31}$ & $^{***}$ & $\mathbf{0.66}$ & $^{***}$ & $\mathbf{0.50}$ &$^{***}$ \\
			\midrule
			$R^{2}$ & $\mathbf{0.14}$ &  & $\mathbf{0.12}$ & & $\mathbf{0.06}$ & \\
			
			\bottomrule
		\end{tabular}
		\caption{The effects of shot ``genderness''.} 
	\end{subtable}	
	\caption{OLS regressions predicting success, controlling for skill (a) and image (b) ``genderness''. (\textit{Note:}  $^{*}p < 0.05$; $^{**}p < 0.01$; $^{***}p < 0.001$)}
	\label{tab:success_maleness}
\end{table*}

\subsection{``Genderness'' of Skills and Products}

In this section we investigate whether women and men have different specializations and create identifiably different products. If so, differences in outcome may come from differences in the taste or size of audiences. For example we observe that women are significantly more likely to list ``copywriting'' as a personal skill. If the audience for copywriting-related work is smaller, or even behaves differently, this may explain differences in outcome.

We implement this idea of ``genderness" of user production at two levels. At the user level we quantify the extent to which skills are listed by males and females. At the shot level we train a neural network to identify shots made by males and females visually. We augment that model with data from the tags users give their images.

\para{Skills}
We calculate the maleness of skills using a log likelihood ratio $L(skill,gender)$:
$$ L(skill,gender) = \log\left(\dfrac{P(skill | gender)}{P(skill)}\right) $$

We test for significance by shuffling the gender of users and calculating a 90\% confidence interval. We adopt this shuffling approach in order to preserve the co-occurrence of skills. Out of 150 skills\footnote{We processed the skills using standard text matching and deduplication techniques.} listed by at least 10 users, 38 skills are deemed significantly male-dominated, while 15 are considered significantly female-dominated. We share the most significantly male and female skills in Table \ref{tab:skills_common}.

We test the impact of this gender difference by adding two binary variables to our original models: whether the user lists a skill categorized as male-dominated, and whether the user lists a skill categorized as female-dominated. The first set of regressions shown in Table~\ref{tab:success_maleness} indicate a penalty for users listing female-dominated skills across all measures of success. Users listing male skills receive more views, suggesting that women who list such skills get similar views to their male counterparts.

\para{Images and tags}
In this subsection, we build a classifier to predict whether a shot has been made by a male or female user. We consider two aspects of the product:  the visual content of the image itself and the tags listed by the user. 

As an image classification problem, identification of the gender of the author is clearly distinguished from typical object 
detection classification tasks in which visual content relates to class labels (e.g. is there a cat in the image?). It also differs substantially from the more refined visual concept detection problems \cite{huiskes2010new} like detecting calming or frightening moods in images. Feed-forward deep neural networks are widely used in various image related tasks such as image captioning \cite{vinyals2016show} and are especially popular for object classification \cite{krizhevsky2012imagenet,he2015deep,szegedy2015rethinking} and object localization \cite{girshick2015fast,ren2015faster}. Due the complexity of the latest models (often containing more than 10 million parameters) the training phase demands a lot of data and computational power. Additionally, almost all feedforward-based discriminative networks have low-dimensional flat layers which can be used as a representation. This led us to use an already tuned model. We chose one of the state-of-the-art feed-forward networks, the Inception v3 \cite{szegedy2015rethinking}. Our choice was mainly based on the quality implementation of the specific model in Tensorflow\footnote{\url{https://github.com/tensorflow/models/tree/master/inception}} and the low dimensional representation (2048) of the images before the final, discriminative layer. To sum up, the neural network generates feature vectors from images, encoding visual data in a way amenable to analysis with a classifier.

Authors tag their shots to help users search for and interpret their work. We consider tags as image related annotations and hence as extra information to aid classification. The sparsity of the tag occurrences keep us from using more complex frequency based textual feature generation techniques (e.g. TF-IDF or Okapi BM25 \cite{robertson1976relevance}), thus we extracted the raw image tags and left out both infrequent and author specific tags. The resulting 10,000 dimensional, sparse representation was normalized and combined with the visual representation of the Inception model into a 12,000 dimensional feature space. 

\begin{table}[t]
	\centering
	\scriptsize
	\begin{tabular}{llll}
		\toprule
		\textbf{Model}	&\textbf{Modality} 	&\textbf{AUC} \\ \hline
		Logistic Regression (LR) & Tags & 0.70 \\
		Inception v3 + LR & Visual & 0.62 \\
		Inception v3 + LR & Mixed & 0.72 \\ 
		\bottomrule
	\end{tabular}
	\caption{ROC AUC measurements of gender classification via visual and textual 
		content.}
	\label{tab:vistag_cf}
\end{table}

\begin{figure}[t]
	\centering \includegraphics[width=0.4\textwidth]
	{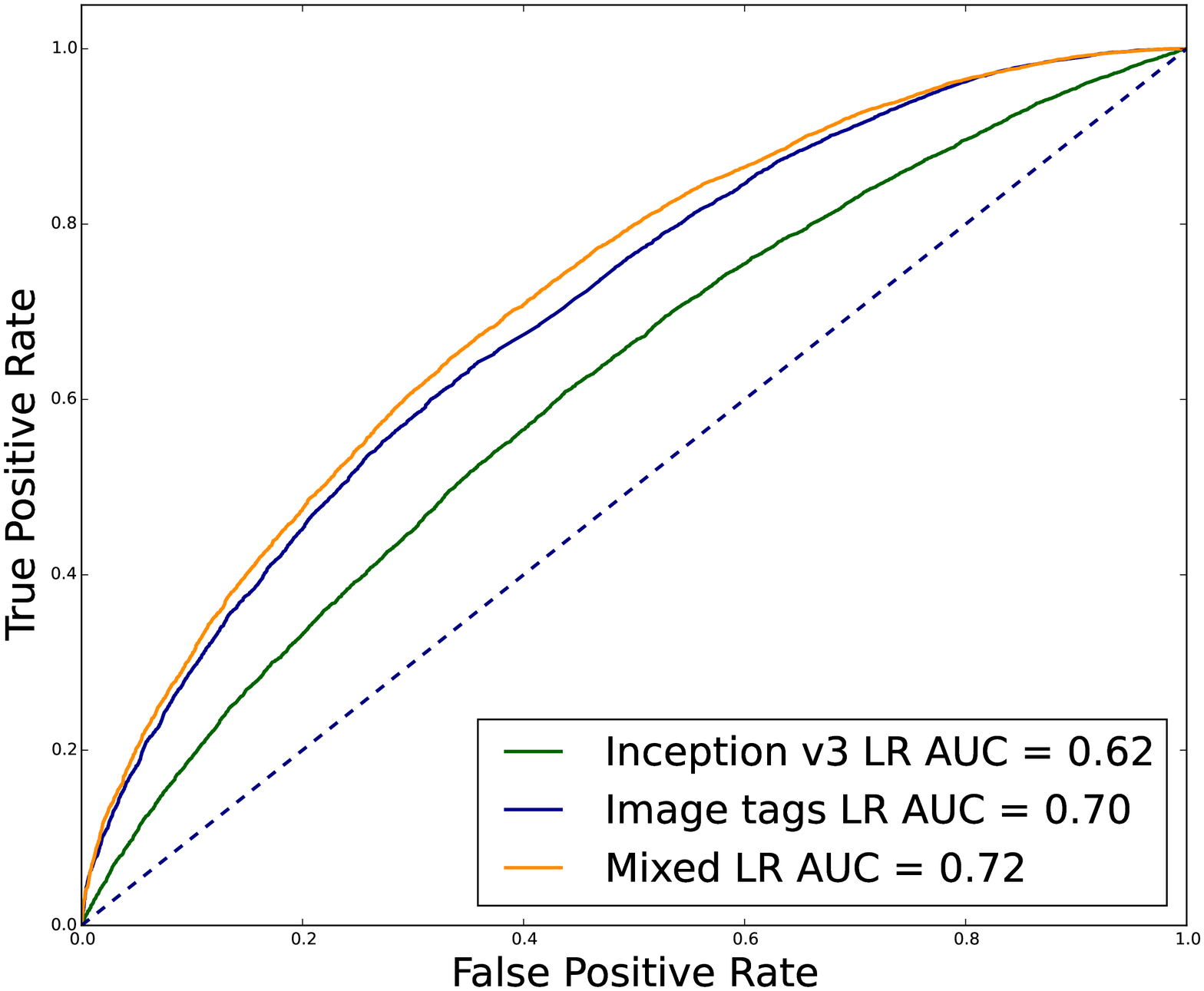}
	\vspace{-1em}
	\caption{ROC curves of the visual, textual and multimodal image classifiers.}
	\label{fig:roc}
\end{figure}

We trained three regularized logistic regression models on the image features, the tag features, and all features combined, respectively. We used ten-fold cross-validation (CV)  \cite{kohavi1995study} and evaluated the results of the classifiers using the area under the curve (AUC) of the receiver operating characteristic (ROC) averaged across the held-out folds. Our choice of ten-fold CV was driven by the findings in \cite{ambroise2002selection}, in which the authors recommended
ten-fold cross-validation instead of leave-one-out. The low variance in AUC across the folds for all three modalities (0.01 for the visual-only, 0.006 for image tags and 0.007 for multimodal on average) suggests the results are robust to overfitting. Table~\ref{tab:vistag_cf} shows that the combined multimodal classification model performed the best, while the image tags outperform the visual model. We plot the ROC curves in Figure~\ref{fig:roc}.

\begin{figure}[t]
	\centering \includegraphics[width=0.4\textwidth]
	{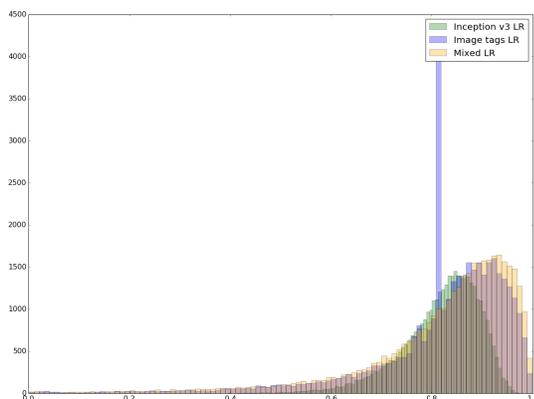}
	\vspace{-1em}
	\caption{The distributions of the hold-out fold predictions for image maleness for the three classifiers. 1 indicates an image with highly male features. Note that the addition of the image features greatly smoothes the distribution of image scores compared to the tag-only model.}
	\label{fig:pred_distr}
\end{figure}

We use the score of the multimodal classifier to assign a gender score to each image. We plot the distributions of the classifier outputs in Figure~\ref{fig:pred_distr}, noting that the while the increase in AUC of the multimodal classifier over the tag-only classifier is relatively small, it greatly smoothes out the predictions of the classifier. The tag-only classifier has difficulty overcoming the sparsity of the tag space. 

We incorporate this new shot level measure of genderness into our regressions by considering a user's average shot maleness. The second set of regressions in Table~\ref{tab:success_maleness} show a significant positive effect of shot maleness on all outcome variables. Moreover, when controlling for this measure the gender of the user is no longer significant for any of the outcomes. Mirroring our findings with the genderness of skills, it seems that authors creating more male images, regardless of their gender, are receiving better outcomes.

\para{Summary}
In this subsection we have presented two ways of quantifying the genderness of a user's output. Both the genderness of a user's skillset and their outputs as defined by our measures have significant relationships with outcomes, indicating that, at least to some extent, differences in gender outcomes on Dribbble are the result of differences in production.

\subsection{Social Behavior and Network Position}

\begin{table*}[t!]
	\scriptsize
	\begin{subtable}[t!]{0.48\textwidth}
		\centering
		\begin{tabular}{lr@{}lr@{}lr@{}l} 
			\toprule
			User Averages (log): & \multicolumn{2}{c}{Views} & \multicolumn{2}{c}{Likes}  & \multicolumn{2}{c}{Responses} \\
			\midrule
			Log(Number of Shots) & $\mathbf{-0.55}$ & $^{***}$ & $\mathbf{-0.50}$ & $^{***}$ & $\mathbf{-0.41}$&$^{***}$\\
			Log(Age of Account) & $\mathbf{-0.07}$ &  $^{**}$& $\mathbf{-0.22}$ & $^{***}$ & $-0.01$ & \\
			Leadership Word in Bio & $-0.04$ &  & $0.001$ & & $0.03$ & \\
			Pro-badge & $0.03$ & & $0.01$ &  &$-0.01$ & \\
			Log(Team Size) & $\mathbf{0.24}$ & $^{***}$ & $\mathbf{0.19}$ & $^{***}$ & $\mathbf{0.11}$ & $^{***}$\\
			Is Male & $0.04$ &  & $-0.04$ &  & $-0.03$ & \\
			Log(Follower Count) & $\mathbf{0.72}$ & $^{***}$ & $\mathbf{0.69}$ & $^{***}$ & $\mathbf{0.49}$ &$^{***}$ \\
			Twitter Reciprocity & $-0.04$ &  & $-0.06$ & & -0.07 &  \\
			Twitter Ego Density & $\mathbf{0.26}$ & $^{***}$ & $\mathbf{0.34}$ & $^{***}$ & $\mathbf{0.27}$ &$^{***}$ \\
			
			\midrule
			
			$R^{2}$ & $\mathbf{0.67}$ &  & $\mathbf{0.67}$ & & $\mathbf{0.50}$ & \\
			
			\bottomrule
		\end{tabular}
		\caption{The effects of network structure on success.}
		\label{tab:success_followers}
	\end{subtable}	
	\begin{subtable}[t!]{0.48\textwidth}
		\centering
		\begin{tabular}{lr@{}lr@{}} 
			\toprule
			User Averages (log): & \multicolumn{2}{c}{Number of Followers}  \\
			\midrule
			Log(Number of Shots) & $\mathbf{0.93}$ & $^{***}$ \\
			Log(Age of Account) & $\mathbf{0.16}$ &  $^{***}$\\
			Leadership Word in Bio & $\mathbf{0.13}$ & $^{**} $  \\
			Pro-badge & $\mathbf{0.66}$ & $^{***}$ \\
			Log(Team Size) & $\mathbf{0.21}$ & $^{***}$ \\
			Is Male & $\mathbf{0.21 }$& $^{***} $\\
			\midrule
			$R^{2}$ & $\mathbf{0.58}$ &  \\
			\bottomrule
		\end{tabular}
		\caption{The effects of user characteristics on follower count.}
		\label{tab:ols_followers}
	\end{subtable}	
	\caption{OLS regressions predicting success and number of followers. (\textit{Note:}  $^{*}p < 0.05$; $^{**}p < 0.01$; $^{***}p < 0.001$)}
\end{table*}

The most important way for Dribbble users to get informed about new work of others is by following them on the site. Indeed there are two ways a viewer may discover a shot: either the viewer follows the user and the user's work is then added to the viewers chronological feed, or the viewer searches for shots by keyword. The former case is clearly a more dependable source of views, likes, and responses. Thus differences between men and women with regard to success on Dribbble may be partly explained by gender differences in follower networks. Not only the sheer size of the networks (number of followers) may matter for user success, but also network structure: people with more cohesive networks may be more efficient in reaching their audience with new work.

Unfortunately, we do not have access to the full follower network on Dribbble, but we can use the Twitter following network among users as a proxy. This choice is justified by the predictive power of the Dribbble user follower counts and the significant correlation ($\rho =.49$) between Dribbble follower count and Twitter follower count. We therefore consider the effect of two features derived from the Twitter network: the reciprocity of ties of a user and the density of a user's ego network.\footnote{The reciprocity of a user in a directed network is defined as the ratio of mutual ties to total ties adjacent to a user. A user's ego network density is defined as number of possible connections between the nodes that the user follows.} In the models described in Table~\ref{tab:success_followers} we see that the number of followers a user has is a very strong predictor of success. The density of a user's ego network predicts success as well and the $R^{2}$ value of all three models increases drastically over that of previous models. Once controlling for these terms, gender becomes insignificant, yet Table~\ref{tab:ols_followers} shows that when we use gender to predict the number of followers a user has, we find that males have an advantage despite several controls. This suggests that gender differences in social network position may indeed lie behind success discrepancies between men and women.


\para{Modeling Social Ties} To be able to distinguish between the impact of multiple factors shaping the network structure of men and women, we use Exponential Random Graph Models (ERGMs) on the observed Twitter network. ERGMs \cite{lusheretal2012} are a multivariate statistical network model family designed exactly for quantifying the contribution of local network configurations (e.g. reciprocated ties, closed triangles) to the global structure of an observed social network, taking into account the inherent dependencies between ties in the network. We estimate ERGMs using the ``statnet'' package implemented in R \cite{handcock:statnet,hunter2007curved}.

Table~\ref{table:ergm_results} summarizes our models and findings. The first model includes basic structural effects (such as reciprocity and transitivity) and gender-related sender, receiver, and similarity effects. The second model contains additional effects describing gender differences in following reciprocity and transitivity. The structural effects draw a similar picture of the network in both models: the following network is characterized by positive tendencies for reciprocity and transitivity, while in- and out-degrees, the number of ties received and sent by users, are relatively evenly distributed (as suggested by the negative parameters for degree effects). 

With regards to gender differences, our two models suggest slightly different patterns. In the first model, men are significantly less likely to follow other users than women, while same gender ties are more likely to exist than cross-gender ties. However, the second model shows that men are more popular (more likely to receive ties) than women, with a similar tendency for same-sex ties. In this latter model, we also included three effects to further study gender differences, which show that reciprocity is less likely between two men and more likely between two women as compared to mixed gender dyads. Finally, tendencies for transitivity, and thereby clustering, do not seem to differ in same gender triads from mixed gender triads.

\begin{table}[t!]
	\centering
	\scriptsize
	\begin{tabular}{lr@{}lr@{}l} 
		\toprule
		Following ties (log odds) & & \\ 
		\midrule
		Edges (Intercept) & $\mathbf{-5.263}$ & $^{***}$ & $\mathbf{-5.682}$ & $^{***}$ \\ 
		Reciprocity (Mutual) & $\mathbf{3.110}$ & $^{***}$  & $\mathbf{3.449}$ & $^{***}$  \\ 
		Transitivity (GWESP, $\sqrt{}$)  & $\mathbf{0.325}$& $^{***}$ & $\mathbf{0.339}$& $^{***}$ \\
		Indegree (GWIn, $\sqrt{}$) & $\mathbf{-3.117}$ & $^{***}$ & $\mathbf{-3.126}$ & $^{***}$ \\  
		Outdegree (GWOut, $\sqrt{}$) & $\mathbf{-2.039}$ & $^{***}$ & $\mathbf{-2.051}$ & $^{***}$ \\ 
		\hline \\[-1.8ex]
		Sender Sex (Male=1) & $\mathbf{-0.169}$ & $^{***}$ & $-0.037$ & \\ 
		Receiver Sex (Male=1) & 0.012 &  & $\mathbf{0.252}$ & $^{***}$ \\
		Same Sex & $\mathbf{0.191}$ & $^{***}$ & $\mathbf{0.234}$ & $^{***}$\\   
		\hline \\[-1.8ex]
		Reciprocity: Both Female  & -- & & $\mathbf{0.367}$ & ${^+}$  \\
		Reciprocity: Both Male & -- & & $\mathbf{-0.641}$ & $^{***}$ \\
		Transitivity: Same Sex & -- & & $\mathbf{-0.000}$ & \\
		\hline \\[-1.8ex]  
		\# of Actors & 3,765 \\ 
		\bottomrule
	\end{tabular} 
	\caption{Exponential Random Graph Models explaining Twitter following networks of users. (\textit{Note:}  $^{+}p < 0.1$; $^{*}p < 0.05$; $^{**}p < 0.01$; $^{***}p < 0.001$)}
	\label{table:ergm_results}
\end{table}

The two models together suggest that although men appear to have a larger number of followers than women, female users of Dribbble tend to have a more cohesive network than men. This tendency is expressed in the first model by the negative sender and positive same sex effect: following between users of the same sex is more likely than between those of different sex, but ties between men are less likely to exist than ties between women because of the lower activity of men. This finding is put in a different light in the second model which suggests that while men attract more followers than women, dyads between women are more likely to be reciprocal. Regarding clustering, we do not find evidence that same-gender triads are more likely to exist than mixed-gender triads, pointing out that gender effects may have a more important role on the level of dyads (homophily, reciprocity) than on the level of groups (transitivity).

\para{Strength of Social Ties.} As a final test of our theory connecting social network position to Dribbble outcomes, we consider the efficiency of a user, defined as the average number of likes per view he or she receives, as another dependent variable. We argue this measures how invested the followers of a user are in his or her work, and how effectively the user broadcasts his or her work. Indeed, as seen in Table~\ref{tab:efficiency}, women have better outcomes according to this measure. We suggest that this is another signal that men and women have structural differences in their social networks and how they navigate them.

\begin{table}[t!]
	\scriptsize
	\centering
	\begin{tabular}{lr@{}lr@{}} 
		\toprule
		User Average: & \multicolumn{2}{c}{Likes per View}  \\
		\midrule
		Log(Number of Shots) & $\mathbf{0.14}$ & $^{***}$ \\
		Log(Age of Account) & $\mathbf{-0.74}$ &  $^{***}$\\
		Leadership Word in Bio & $0.13$ &   \\
		Pro-badge & $-0.08$ & \\
		Log(Team Size) & $\mathbf{-0.19}$ & $^{***}$ \\
		Is Male & $\mathbf{-0.39 }$& $^{***} $\\
		Twitter Reciprocity & -0.14&\\
		Twitter Ego Density & 0.21 & \\
		
		\midrule
		
		$R^{2}$ & $\mathbf{0.17}$ &  \\
		
		\bottomrule
	\end{tabular}
	
	\caption{OLS regression predicting likes per view, controlling for user follower count. (\textit{Note:}  $^{*}p < 0.05$; $^{**}p < 0.01$; $^{***}p < 0.001$)}
	\label{tab:efficiency}
\end{table}

\section{Discussion}

In this paper, we studied the gender differences of individual success in an online community for graphic designers. Our first results showed that men tend to be more successful than women even when controlling for differences including activity, tenure, status markers, and investment, though none of the models created to test these relationships could explain much variation in outcome. Consequently, we turned our attention to how male and female designers may have different audiences for their products or distinct social networks structures, to help better understand the observed results.  

When looking at the skills and images users create, we find that there is a significant subset of the top skills that are highly male or female dominated. Using the images and their descriptive tags we are able to predict the gender of the creator with an AUC of 0.72. Furthermore skill and image ``genderness'' explain some of the gender differences behind success: women with male skills have similar success rates to males and vice versa. This suggests that part of the gender gap comes from different production patterns.

Finally, analysis on the social network underlying Dribbble shows that the number of followers a user has in the community captures a large part of the variation in individual outcomes. Men produce more work and have more followers, thus they receive more views, likes and responses overall. However, a closer look at the social network structure shows that while men have more followers, women tend to be part of smaller, more densely knit clusters with more reciprocal ties and thus are able to turn their image views into likes and responses more frequently relative to men. As a limitation of our research, we acknowledge that the Twitter follower network might not correspond precisely with the Dribbble social network, and suggest that there is value in extending the data collection effort to the Dribbble network.

Our results demonstrate that there can be multiple potential sources of gender inequalities in online markets. However, this study only puts forth a few simple mechanisms that may be driving these biases. In reality, individual characteristics and social structures are dynamically interrelated: users may adjust their production and self-representation to be able to reach larger audiences, and social relations constantly evolve as a result of users trying to adapt to an ever changing environment. What seems clear, however, is that ``gender'' remains one of the key categories around which communities produce their understanding of quality and success.

In addition, the described empirical patterns of success also suggest a few lessons for the design of algorithms to present the work of users. Dribbble incorporates common design elements of today's social and labor market platforms, such as sharing content to social ties, publicly visible feedback/success measures, and content recommendation based on popularity/relevance. For example, if search engine on Dribbble ranks relevant images using views, it indirectly advantages men. If it were to rank images using likes per view, women would be advantaged. This kind of detail is important, given that feedback loops and rich get richer effects can inflate differences in outcomes over time. It certainly merits further investigation by researchers, designers of online platforms, and regulators.

\section{Acknowledgments}
We thank the anonymous reviewers for their extremely helpful comments. We also thank our colleagues Karl Wachs, Zs\'ofia Cz\'em\'an, J\'ulia Koltai, Christo Wilson, and Roberta Sinatra for helpful input and feedback.

\fontsize{9.0pt}{10.0pt}
\selectfont
\bibliographystyle{ieeetr}
\bibliography{biblio}

\begin{thebibliography}{10}

\bibitem{lustig2016algorithmic}
C.~Lustig, K.~Pine, B.~Nardi, L.~Irani, M.~K. Lee, D.~Nafus, and C.~Sandvig,
  ``Algorithmic authority: the ethics, politics, and economics of algorithms
  that interpret, decide, and manage,'' in {\em Proceedings of the 2016 CHI
  Conference Extended Abstracts on Human Factors in Computing Systems},
  pp.~1057--1062, ACM, 2016.

\bibitem{lee-chi15}
M.~K. Lee, D.~Kusbit, E.~Metsky, and L.~Dabbish, ``Working with machines: The
  impact of algorithmic and data-driven management on human workers,'' in {\em
  Proceedings of the 33rd Annual ACM Conference on Human Factors in Computing
  Systems}, 2015.

\bibitem{teodoro-2014-cscw}
R.~Teodoro, P.~Ozturk, M.~Naaman, W.~Mason, and J.~Lindqvist, ``The motivations
  and experiences of the on-demand mobile workforce,'' in {\em Proceedings of
  the 17th ACM Conference on Computer Supported Cooperative Work and; Social
  Computing}, 2014.

\bibitem{thebault-2015-cscw}
J.~Thebault-Spieker, L.~G. Terveen, and B.~Hecht, ``Avoiding the south side and
  the suburbs: The geography of mobile crowdsourcing markets,'' in {\em
  Proceedings of the 18th ACM Conference on Computer Supported Cooperative Work
  and Social Computing}, 2015.

\bibitem{hannak-2017-cscw}
A.~Hannak, C.~Wagner, D.~Garcia, A.~Mislove, M.~Strohmaier, and C.~Wilson,
  ``{Bias in Online Freelance Marketplaces: Evidence from TaskRabbit and
  Fiverr},'' in {\em {20th ACM Conference on Computer-Supported Cooperative
  Work and Social Computing (CSCW 2017)}}, (Portland, OR), February 2017.

\bibitem{vasilescu2013gender}
B.~Vasilescu, A.~Capiluppi, and A.~Serebrenik, ``Gender, representation and
  online participation: A quantitative study,'' {\em Interacting with
  Computers}, p.~iwt047, 2013.

\bibitem{de2015game}
M.~De~Vaan, D.~Stark, and B.~Vedres, ``Game changer: The topology of
  creativity1,'' {\em American Journal of Sociology}, vol.~120, no.~4,
  pp.~1144--1194, 2015.

\bibitem{angristlang2004}
J.~D. Angrist and K.~Lang, ``Does school integration generate peer effects?
  evidence from boston's metco program,'' {\em American Economic Review},
  pp.~1613--1634, 2004.

\bibitem{curcio2006sleep}
G.~Curcio, M.~Ferrara, and L.~De~Gennaro, ``Sleep loss, learning capacity and
  academic performance,'' {\em Sleep medicine reviews}, vol.~10, no.~5,
  pp.~323--337, 2006.

\bibitem{coleman1994}
J.~S. Coleman, {\em Foundations of Social Theory}.
\newblock Harvard University Press, 1994.

\bibitem{feld1982}
S.~L. Feld, ``Social structural determinants of similarity among associates,''
  {\em American Sociological Review}, vol.~47, no.~6, pp.~797--801, 1982.

\bibitem{mcphersonetal2001}
M.~McPherson, L.~Smith-Lovin, and J.~M. Cook, ``Birds of a feather: Homophily
  in social networks,'' {\em Annual Review of Sociology}, vol.~27,
  pp.~415--444, 2001.

\bibitem{heider1946}
F.~Heider, ``Attitudes and cognitive organization,'' {\em The Journal of
  Psychology}, vol.~21, no.~1, pp.~107--112, 1946.

\bibitem{cartwrightharary1956}
D.~Cartwright and F.~Harary, ``Structural balance: a generalization of heider's
  theory.,'' {\em Psychological review}, vol.~63, no.~5, p.~277, 1956.

\bibitem{davis1970}
J.~A. Davis, ``Clustering and hierarchy in interpersonal relations: Testing two
  graph theoretical models on 742 sociomatrices,'' {\em American Sociological
  Review}, vol.~35, no.~5, pp.~843--851, 1970.

\bibitem{veenstradijkstra2011}
R.~Veenstra and J.~K. Dijkstra, ``Transformations in adolescent peer
  networks,'' {\em Relationship pathways: From adolescence to young adulthood},
  pp.~135--154, 2011.

\bibitem{marsdenfriedkin1993}
P.~V. Marsden and N.~E. Friedkin, ``Network studies of social influence,'' {\em
  Sociological Methods \& Research}, vol.~22, no.~1, pp.~127--151, 1993.

\bibitem{turner1991}
J.~C. Turner, {\em Social influence}.
\newblock Milton Keynes, UK: Open University Press, 1991.

\bibitem{petersen2014reputation}
A.~M. Petersen, S.~Fortunato, R.~K. Pan, K.~Kaski, O.~Penner, A.~Rungi,
  M.~Riccaboni, H.~E. Stanley, and F.~Pammolli, ``Reputation and impact in
  academic careers,'' {\em Proceedings of the National Academy of Sciences},
  vol.~111, no.~43, pp.~15316--15321, 2014.

\bibitem{petersen2015}
A.~M. Petersen, ``Quantifying the impact of weak, strong, and super ties in
  scientific careers,'' {\em Proceedings of the National Academy of Sciences},
  vol.~112, no.~34, pp.~E4671--E4680, 2015.

\bibitem{Sarigoel2014}
E.~Sarigöl, R.~Pfitzner, I.~Scholtes, A.~Garas, and F.~Schweitzer, ``Predicting
  scientific success based on coauthorship networks,'' {\em EPJ Data Science},
  2014.

\bibitem{servia2015evolution}
S.~Servia-Rodr{\'\i}guez, A.~Noulas, C.~Mascolo, A.~Fern{\'a}ndez-Vilas, and
  R.~P. D{\'\i}az-Redondo, ``The evolution of your success lies at the centre
  of your co-authorship network,'' {\em PloS one}, vol.~10, no.~3, p.~e0114302,
  2015.

\bibitem{jacobs1996gender}
J.~A. Jacobs, ``Gender inequality and higher education,'' {\em Annual Review of
  Sociology}, pp.~153--185, 1996.

\bibitem{pager-2008-sociology}
D.~Pager and H.~Shepherd, ``The sociology of discrimination: Racial
  discrimination in employment, housing, credit, and consumer markets,'' {\em
  Annual review of sociology}, vol.~34, p.~181, 2008.

\bibitem{clauset2015systematic}
A.~Clauset, S.~Arbesman, and D.~B. Larremore, ``Systematic inequality and
  hierarchy in faculty hiring networks,'' {\em Science Advances}, vol.~1,
  no.~1, p.~e1400005, 2015.

\bibitem{lee2013bias}
C.~J. Lee, C.~R. Sugimoto, G.~Zhang, and B.~Cronin, ``Bias in peer review,''
  {\em Journal of the American Society for Information Science and Technology},
  vol.~64, no.~1, pp.~2--17, 2013.

\bibitem{sandvig2014auditing}
C.~Sandvig, K.~Hamilton, K.~Karahalios, and C.~Langbort, ``Auditing algorithms:
  Research methods for detecting discrimination on internet platforms,'' in
  {\em Proceedings of ``Data and Discrimination: Converting Critical Concerns
  into Productive Inquiry'', a preconference at the 64th Annual Meeting of the
  International Communication Association}, 2014.

\bibitem{robinson2015digital}
L.~Robinson, S.~R. Cotten, H.~Ono, A.~Quan-Haase, G.~Mesch, W.~Chen, J.~Schulz,
  T.~M. Hale, and M.~J. Stern, ``Digital inequalities and why they matter,''
  {\em Information, Communication \& Society}, vol.~18, no.~5, pp.~569--582,
  2015.

\bibitem{tang_dasfaa11}
C.~Tang, K.~W. Ross, N.~Saxena, and R.~Chen, ``What's in a name: A study of
  names, gender inference, and gender behavior in facebook.,'' in {\em DASFAA
  Workshops}, 2011.

\bibitem{lee-2015-chi}
M.~K. Lee, D.~Kusbit, E.~Metsky, and L.~Dabbish, ``Working with machines: The
  impact of algorithmic and data-driven management on human workers,'' in {\em
  Proceedings of the 33rd Annual ACM Conference on Human Factors in Computing
  Systems}, 2015.

\bibitem{adali2010measuring}
S.~Adali, R.~Escriva, M.~K. Goldberg, M.~Hayvanovych, M.~Magdon-Ismail, B.~K.
  Szymanski, W.~A. Wallace, and G.~Williams, ``Measuring behavioral trust in
  social networks,'' in {\em Intelligence and Security Informatics (ISI), 2010
  IEEE International Conference on}, pp.~150--152, IEEE, 2010.

\bibitem{hannak2014get}
A.~Hannak, D.~Margolin, B.~Keegan, and I.~Weber, ``Get back! you don't know me
  like that: The social mediation of fact checking interventions in twitter
  conversations.,'' in {\em ICWSM}, 2014.

\bibitem{munger2016tweetment}
K.~Munger, ``Tweetment effects on the tweeted: Experimentally reducing racist
  harassment,'' 2016.

\bibitem{ajrouch2005social}
K.~J. Ajrouch, A.~Y. Blandon, and T.~C. Antonucci, ``Social networks among men
  and women: The effects of age and socioeconomic status,'' {\em The Journals
  of Gerontology Series B: Psychological Sciences and Social Sciences},
  vol.~60, no.~6, pp.~S311--S317, 2005.

\bibitem{NBERw22776}
Y.~Ge, C.~R. Knittel, D.~MacKenzie, and S.~Zoepf, ``Racial and gender
  discrimination in transportation network companies,'' Working Paper 22776,
  National Bureau of Economic Research, October 2016.

\bibitem{Fradkin:2015:BRO:2764468.2764528}
A.~Fradkin, E.~Grewal, D.~Holtz, and M.~Pearson, ``Bias and reciprocity in
  online reviews: Evidence from field experiments on airbnb,'' in {\em
  Proceedings of the Sixteenth ACM Conference on Economics and Computation},
  2015.

\bibitem{pan-2007-trust}
B.~Pan, H.~Hembrooke, T.~Joachims, L.~Lorigo, G.~Gay, and L.~Granka, ``{In
  Google We Trust: Users' Decisions on Rank, Position, and Relevance},'' {\em
  J. Comp. Med. Comm.}, vol.~12, no.~3, 2007.

\bibitem{babyname}
``Baby names from social security card applications-national level data.''
  data.gov, 2016.
\newblock
  \url{https://catalog.data.gov/dataset/baby-names-from-social-security-card-applications-national-level-data}.

\bibitem{ruths-aaai13}
W.~Liu and D.~Ruths, ``What's in a name? using first names as features for
  gender inference in twitter,'' 2013.

\bibitem{karimi-www16}
F.~Karimi, C.~Wagner, F.~Lemmerich, M.~Jadidi, and M.~Strohmaier, ``Inferring
  gender from names on the web: A comparative evaluation of gender detection
  methods,'' WWW '16 Companion, 2016.

\bibitem{cfpd_race}
``Using publicly available information to proxy for unidentified race and
  ethnicity.'' Consumer Financial Protection Bureau.
\newblock
  \url{https://www2.census.gov/topics/genealogy/2000surnames/surnames.pdf}.

\bibitem{fiscella-hsr06}
K.~Fiscella and A.~M. Fremont, ``Use of geocoding and surname analysis to
  estimate race and ethnicity,'' Health Serv Res, 2006.

\bibitem{huiskes2010new}
M.~J. Huiskes, B.~Thomee, and M.~S. Lew, ``New trends and ideas in visual
  concept detection: the mir flickr retrieval evaluation initiative,'' in {\em
  Proceedings of the international conference on Multimedia information
  retrieval}, pp.~527--536, ACM, 2010.

\bibitem{vinyals2016show}
O.~Vinyals, A.~Toshev, S.~Bengio, and D.~Erhan, ``Show and tell: Lessons
  learned from the 2015 mscoco image captioning challenge,'' {\em IEEE
  transactions on pattern analysis and machine intelligence}, 2016.

\bibitem{krizhevsky2012imagenet}
A.~Krizhevsky, I.~Sutskever, and G.~E. Hinton, ``Imagenet classification with
  deep convolutional neural networks,'' in {\em Advances in neural information
  processing systems}, pp.~1097--1105, 2012.

\bibitem{he2015deep}
K.~He, X.~Zhang, S.~Ren, and J.~Sun, ``Deep residual learning for image
  recognition,'' {\em arXiv preprint arXiv:1512.03385}, 2015.

\bibitem{szegedy2015rethinking}
C.~Szegedy, V.~Vanhoucke, S.~Ioffe, J.~Shlens, and Z.~Wojna, ``Rethinking the
  inception architecture for computer vision,'' {\em arXiv peprint
  arXiv:1512.00567}, 2015.

\bibitem{girshick2015fast}
R.~Girshick, ``Fast r-cnn,'' in {\em Proceedings of the IEEE International
  Conference on Computer Vision}, pp.~1440--1448, 2015.

\bibitem{ren2015faster}
S.~Ren, K.~He, R.~Girshick, and J.~Sun, ``Faster r-cnn: Towards real-time
  object detection with region proposal networks,'' in {\em Neural Information
  Processing Systems (NIPS), 2015}, 2015.

\bibitem{robertson1976relevance}
S.~E. Robertson and K.~S. Jones, ``Relevance weighting of search terms,'' {\em
  Journal of the American Society for Information science}, vol.~27, no.~3,
  pp.~129--146, 1976.

\bibitem{kohavi1995study}
R.~Kohavi {\em et~al.}, ``A study of cross-validation and bootstrap for
  accuracy estimation and model selection,'' in {\em Ijcai}, vol.~14,
  pp.~1137--1145, Stanford, CA, 1995.

\bibitem{ambroise2002selection}
C.~Ambroise and G.~J. McLachlan, ``Selection bias in gene extraction on the
  basis of microarray gene-expression data,'' {\em Proceedings of the national
  academy of sciences}, vol.~99, no.~10, pp.~6562--6566, 2002.

\bibitem{lusheretal2012}
D.~Lusher, J.~Koskinen, and G.~Robins, {\em Exponential Random Graph Models for
  Social Networks: Theory, Methods, and Applications}.
\newblock Cambridge University Press, 2012.

\bibitem{handcock:statnet}
M.~S. Handcock, D.~R. Hunter, C.~T. Butts, S.~M. Goodreau, and M.~Morris, {\em
  statnet: Software tools for the Statistical Modeling of Network Data}.
\newblock Seattle, WA, 2003.

\bibitem{hunter2007curved}
D.~R. Hunter, ``Curved exponential family models for social networks,'' {\em
  Social networks}, vol.~29, no.~2, pp.~216--230, 2007.

\end{thebibliography}


\end{document}